\documentclass[aps, prb, reprint,groupedaddress, showpacs]{revtex4-1}
\usepackage{graphicx}
\usepackage{bm}
\usepackage{comment} 

\begin{document}


\title{Photostimulated desorption of Xe from Au(001) surfaces via transient Xe$^{-}$ formation}


\author{Akihiko Ikeda}\email[E-mail: ]{a-ikeda@iis.u-tokyo.ac.jp}
\author{Masuaki Matsumoto}
\author{Shohei Ogura}
\author{Katsuyuki Fukutani}
\email[E-mail: ]{fukutani@iis.u-tokyo.ac.jp}
\author{Tatsuo Okano}\email[E-mail: ]{okano@iis.u-tokyo.ac.jp}
\affiliation{Institute of Industrial Science, The University of Tokyo, Komaba, Meguro-ku, Tokyo 153-8505, Japan}



\date{\today}

\begin{abstract}
Photo-stimulated desorption (PSD) of Xe atoms from the Au(001) surface in thermal and nonthermal regimes was investigated by the time-of-flight measurement at photon energies of 6.4 and 2.3 eV. Xe was desorbed in a thermal way at high laser fluence, which was in good agreement with theoretical simulations. At a low laser fluence, on the other hand, desorption was induced only at a photon energy of 6.4 eV by a non-thermal one-photon process. We argue that the nonthermal PSD occurs via transient formation of Xe$ ^ { - } $ on Au(001). The lifetime of Xe$ ^ { - } $ is estimated to be $\sim$15 fs with a classical model calculation.  Whereas the electron affinity of Xe is negative in the isolated state, it is stabilized by the metal proximity effect.
\end{abstract}




\maketitle

\section{Introduction}
Photo-stimulated processes at solid surfaces have been a topic of extensive studies because they allow us to control adsorbates in either thermal or nonthermal ways.\cite{Ho1, Chuang, Fukutani2003} Laser-induced thermal desorption (LITD) was investigated in detail for the systems of CO/Fe(110) (Ref.~\onlinecite{Wedler}) and Xe/Cu,\cite{Hussla} and has been successfully applied to the studies of surface diffusion combined with low-energy electron microscopy\cite{Yim} or with scanning tunneling microscopy.\cite{Schwalb} Nonthermal photostimulated phenomena, on the other hand, provide us with pathways that are nonaccessible in a thermal process. The nonthermal photostimulated desorption (PSD) of rare gas atoms from metal surfaces has been investigated using photons of two energy regions: At $h\nu>$ 7 eV, the excitonic or ionic excitation of the mono and multi-layers of Ar and Kr induces desorption,\cite{Feulner1987} while infrared light at $h\nu<$ 1 eV causes the direct excitation of the vibrational mode in the physisorption well to a continuum state.\cite{Pearlstine}

The non-thermal PSD of Xe/metal using 1$-$7 eV photons has been considered not to occur. Generally, the mechanism of nonthermal PSD using photons of 1$-$7 eV is understood in terms of formation of the transient negative ion (TNI)\cite{Richter} and the Antoniewicz model,\cite{Antoniewicz} where a substrate conduction electron is photoexcited to the adsorbate affinity level. The ground state Xe in the gas phase does not bind an electron stably,\cite{Buckman, Nicolaides, Bae, Hird} which has been confirmed both theoretically and experimentally with the exception of Ref.~\onlinecite{Haberland}. Xe atoms physisorb on a metal surface. Physisorption is assumed to occur with little influence on the electronic states. Hence, it has been anticipated that the PSD of Xe/metal via TNI is absent. Condensed Xe, however, has been reported to have modified electronic states compared with the isolated ones due to hybridization with the orbitals of neighboring atoms. It is known that it takes 0.5 eV to remove an excess electron from the bulk Xe,\cite{Schwentner1975} and also that the ground state Xe$_{N}$ clusters with $N>6$ stably bind an electron,\cite{Haberland} indicating that the electron affinity level of Xe is shifted downward or broadened depending on the phase of Xe. In this sense, adsorption of Xe onto metal surfaces may well result in a shift and/or broadening or even narrowing of its affinity level by hybridization of the unoccupied orbitals with the substrate electronic states, as is predicted by theoretical studies.\cite{Nordlander, Silva}

In the present paper, we report an experimental study of LITD and nonthermal PSD of Xe from a Au(001) surface at photon energies of 6.4 and 2.3 eV. At a high laser fluence, Xe desorption was thermally induced at both photon energies, which is in good agreement with theoretical calculations. At a low laser fluence, on the other hand, Xe desorption was induced nonthermally by 6.4 eV photons as a one-photon process, whereas little desorption was observed with 2.3 eV photons. We argue that the nonthermal PSD proceeds with a transient formation of Xe$^{-}$ as a result of the photoexcitation of substrate conduction electrons. A classical model calculation of Xe desorption reproduces both the experimentally observed TOF and nonthermal PSD cross section, assuming a value of the Xe$^{-}$ lifetime to be $\sim$15 fs.

\section{Experiment}
A single-crystal disk of Au(001) was mounted on a cold head after being chemically and mechanically polished. The Au(001) surface was cleaned by several cycles of Ar$^{+}$ ion sputtering at 0.5 keV and annealing at 700 K in an ultrahigh vacuum chamber ($p=2.0\times10^{-8}$ Pa). The cleanliness of Au(001) was confirmed by observing the ($5\times20$) reconstructed pattern with LEED (Ref.~\onlinecite{Vanhove}) and no contamination in AES. A Xe monolayer was formed by dosing Xe gas of 3 L (1 L = $1.33\times10^{-4}$ Pa s) to the sample surface at 23 K (Ref.~\onlinecite{Dai}) cooled by a closed-cycle He-compression type refrigerator. The ArF excimer laser (6.4 eV, 8 ns) and the second harmonics of Nd:YAG (yttrium aluminum garnet) laser (2.3 eV, 7 ns) pulses were guided onto the area of 1 mm$^{2}$ on the sample surface with an incidence angle of 25$^{\circ}$ from the surface normal. The desorbing Xe atoms were detected by a quadrupole mass spectrometer (QMS) located in the surface normal direction with a flight distance of 10 cm. The output signal was amplified by a fast current amplifier and recorded with an oscilloscope synchronized with the laser pulse. The QMS was operated in a low-resolution mode, which enabled high-sensitivity detection of Xe atoms.

\section{Results}
\begin{figure}
\begin{center}
\includegraphics[scale=.6]{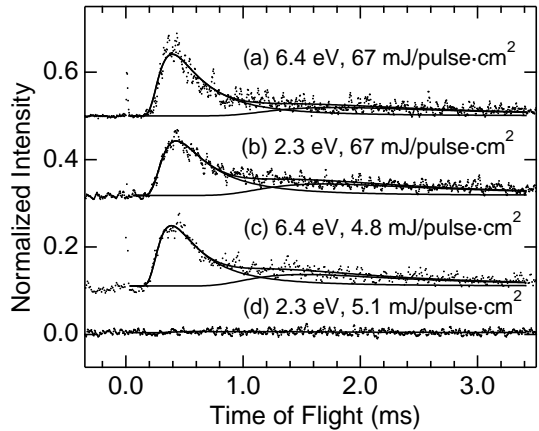} 
\caption{Time-of-flight spectra of desorbing Xe atoms from Au (001) following pulse laser irradiation. On each spectra, the photon energy and the absorbed energy by the sample are denoted. Data of (a), (b) and (c), (d) were recorded with a single pulse and the accumulation of over 120 pulses, respectively.\label{tof}}
\end{center}
\end{figure}

The time-of-flight (TOF) of desorbing Xe atoms upon laser irradiation was recorded at a wide range of laser pulse energy absorbed by the sample ($I_{\mathrm{L}}$) for both 6.4 and 2.3 eV photons. $I_{\mathrm{L}}$ was estimated by taking account of the reflectivity on Au (0.8 for 2.3 eV and 0.2 for 6.4 eV). Figure~\ref{tof} shows typical TOF results. The data reveal a maximum at a TOF of 400 $\mu$s with a tailing feature in the long TOF region. TOF was recorded with only one pulse for both 6.4 and 2.3 eV photons with $I_{\mathrm{L}}>$ 10 mJ/pulse cm$^{2}$ [Figs.~\ref{tof}(a) and \ref{tof}(b)]. With $I_{\mathrm{L}}<$ 10 mJ/pulse cm$^{2}$, on the other hand, TOF was recorded by accumulating over 120 data [Figs.~\ref{tof}(c) and \ref{tof}(d)] because the Xe desorption yield was small. Whereas a substantial desorption yield was observed with 6.4 eV photons as shown in Fig.~\ref{tof}(c), no significant signal was recorded with 2.3 eV photons as shown in Fig.~\ref{tof}(d). Solid curves in Fig.~\ref{tof} are fits to the data with a sum of two Maxwell-Boltzmann (MB) velocity distributions described as $f(v)=A_{1}v^{2}\exp(-mv^{2}/2kT_{\mathrm{D1}})+A_{2}v^{2}\exp(-mv^{2}/2kT_{\mathrm{D2}})$, where $m$ and $k$ are mass of a Xe atom and the Boltzmann constant, respectively, and $A_{i}$ and the translational temperature $T_{\mathrm{D}i}$ are fitting parameters. In the analysis, the form is converted to the flux weighted form. In the following, we discuss only the fast component of the TOF.

\begin{figure}
\begin{center}
\includegraphics[scale=.6]{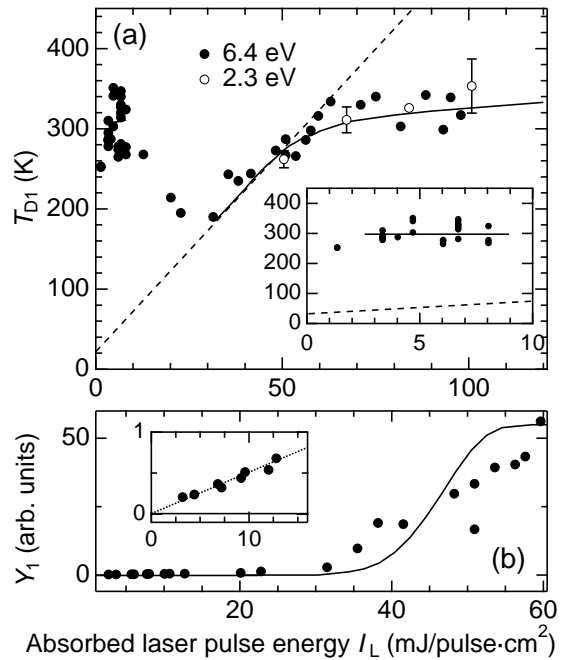} 
\caption{(a) Translational temperature ($T_{\mathrm{D}1}$) of Xe atoms desorbed from Au(001) following laser irradiation as a function of absorbed laser pulse energy ($I_{\mathrm{L}}$). The solid and dashed lines are calculated results of the surface temperature at the maximum desorption rate ($T_{\mathrm{DM}}$) and the maximum surface temperature ($T_{\mathrm{SM}}$), respectively. The inset is a magnification of the small $I_{\mathrm{L}}$ region. (b) Desorption yield of Xe atoms ($Y_{\mathrm{1}}$) following the laser irradiations with 6.4 eV photons as a function of $I_{\mathrm{L}}$. Calculated results of $Y_{\mathrm{1}}$ are shown by the solid curve. The inset is a magnification of the small $I_{\mathrm{L}}$ region. \label{td}}
\end{center}
\end{figure}

Figures~\ref{td}(a) and \ref{td}(b) show $T_{\mathrm{D}1}$ and the Xe desorption yield of the first component ($Y_{\mathrm{1}}$) plotted as a function of $I_{\mathrm{L}}$, respectively. We first focus on the region of $I_{\mathrm{L}}>32$ mJ/pulse cm$^{2}$. In this region, $T_{\mathrm{D}1}$ increases with increasing $I_{\mathrm{L}}$ from about 200 K at 32 mJ/pulse cm$^{2}$ and saturates at about 300 K, for both 6.4 and 2.3 eV photons. In Fig~\ref{td}(b), $Y_{\mathrm{1}}$ shows a sharp increase at 32 mJ/pulse cm$^{2}$.  The behavior of $T_{\mathrm{D}1}$ and $Y_{\mathrm{1}}$ indicates that the Xe desorption is thermally activated with $I_{\mathrm{L}}$ $>32$ mJ/pulse cm$^{2}$.\cite{Wedler, Hussla} For the quantitative analysis, we carried out numerical calculations of the surface temperature ($T_{\mathrm{S}}$) during laser irradiation on the basis of the one-dimensional heat conduction equation.\cite{Hicks} Subsequently, the time evolution of the Xe coverage and the Xe desorption rate (i.e., LITD) was deduced by employing the first-order desorption kinetics assuming the activation energy for desorption of Xe from Au(001) to be 240 meV.\cite{Mcelhiney} By the calculations above, we obtained the surface temperature at maximum Xe desorption rate ($T_{\mathrm{DM}}$) and the maximum surface temperature ($T_{\mathrm{SM}}$) following laser pulse irradiation with $I_{\mathrm{L}}$.

$T_{\mathrm{DM}}$ and $T_{\mathrm{SM}}$ deduced from the calculation are depicted as a function of $I_{\mathrm{L}}$ in Fig.~\ref{td}(a) as dashed and solid curves, respectively. $T_{\mathrm{D1}}$ obtained by the experimental results is in good agreement with $T_{\mathrm{DM}}$, indicating that LITD of Xe is dominant with $I_{\mathrm{L}}>32$ mJ/pulse cm$^{2}$. $T_{\mathrm{DM}}$ deviates from $T_{\mathrm{SM}}$ with $I_{\mathrm{L}}>50$ mJ/pulse cm$^{2}$ because desorption occurs before the surface temperature reaches its maximum.\cite{Wedler} The solid line in Fig.~\ref{td}(b) shows a calculated result of the $Y_{\mathrm{1}}$ by LITD as a function of $I_{\mathrm{L}}$. The calculated $Y_{\mathrm{1}}$ sharply increases at 35 mJ/pulse m$^{2}$ and saturates above 55 mJ/pulse cm$^{2}$. This is in good agreement with the experimental data that the desorption yield exhibits a steep increase at 32 mJ/pulse cm$^{2}$. This thresholdlike behavior is typical of LITD. However, the experimental data in Fig.~\ref{td}(b) monotonously increases in contrast to the saturating behavior of the calculated curve, which may be caused by either spatial inhomogeneity of the desorption laser intensity\cite{Koehler} or coverage dependence of the activation energy for desorption due to the attractive interactions between the adsorbates.\cite{Wedler}

We turn next to the region of $I_{\mathrm{L}}<$ 24 mJ/pulse cm$^{2}$, where desorption of Xe is observed only at 6.4 eV. As can be seen in Fig.~\ref{td}(a), $T_{\mathrm{D}1}$ significantly deviates from the calculated result of $T_{\mathrm{SM}}$ in this region of $I_{\mathrm{L}}$. The inset in Fig.~\ref{td}(a) shows a magnification of $T_{\mathrm{D}1}$ with $I_{\mathrm{L}}<10$ mJ/pulse cm$^{2}$. In this region, $T_{\mathrm{D}1}$ is independent of $I_{\mathrm{L}}$ and constant at 300$\pm$20 K which is much higher than $T_{\mathrm{SM}}$ ($<$100 K). Since in the region of $I_{\mathrm{L}}<24$  mJ/pulse cm$^{2}$ the calculated result of LITD fails to account for the experimental data, other desorption mechanisms should be operative. Especially with $I_{\mathrm{L}}<10$ mJ/pulse cm$^{2}$, the LITD yield of Xe is negligible because $T_{\mathrm{SM}}$ is too low for the thermal activation of Xe desorption. Therefore, only non-thermal PSD of Xe atoms from Au(001) is operative in this region of $I_{\mathrm{L}}$. As shown in the inset of $Y_{\mathrm{1}}$ as a function of $I_{\mathrm{L}}$ in Fig.~\ref{td}(b), $Y_{\mathrm{1}}$ with 6.4 eV photons linearly increases with increasing $I_{\mathrm{L}}$, indicating that the observed nonthermal PSD is a one-photon process. The nonthermal PSD cross section $\sigma_{\mathrm{PSD}}$ was deduced to be 10$^{-21}$$-$10$^{-22}$ cm$^{2}$ by comparing the nonthermal PSD yield with the LITD yield of the Xe monolayer.

\section{Discussion}
We first discuss the initial excitation of the non-thermal PSD of Xe from Au(001) upon irradiation of 6.4 eV photons. As the initial excitation, we argue that the negative ion state of Xe is formed via the photoexcitation of the substrate electron.\cite{Richter} Other excitation pathways can be excluded for the following reasons. The first-excitation energy of Xe from the ground state (5$s^{2}$5$p^{6}$) to the metastable state (5$s^{2}$5$p^{5}$6$s$) is 8.3 eV.\cite{Feulner1987} When Xe is condensed into a two-dimensional layer on a surface, the excitation energy might be modified, as denoted by the surface exciton. The value is, however, reported to be little modified in the monolayer adsorption regime,\cite{Schonhense} suggesting that such excitation is unlikely to occur at 6.4 eV. The first ionization energy is 12.1 eV,\cite{Feulner1984, Moog} which is also unreachable with 6.4 eV photons even though it is reduced due to the image charge effect by $\sim$2.9 eV.\cite{Schonhense} Xe desorption from Ag nanoparticles (AgNP) or Si(001), on the other hand, is also reported to occur via surface-plasmon excitation of AgNP at 2.3$-$4.0 eV photons\cite{Watanabe2007} and localized surface phonon excitation of Si(001) at 1.1$-$6.4 eV photons.\cite{Watanabe2000} Desorption via direct excitation from the bound state to a continuum state took place at a photon energy of lower than 1 eV.\cite{Pearlstine, Rao} All these desorption mechanisms can be ruled out because the nonthermal PSD of Xe was observed only at 6.4 eV and not at 2.3 eV photoirradiation.

\begin{figure}
\begin{center}
\includegraphics[scale=.5]{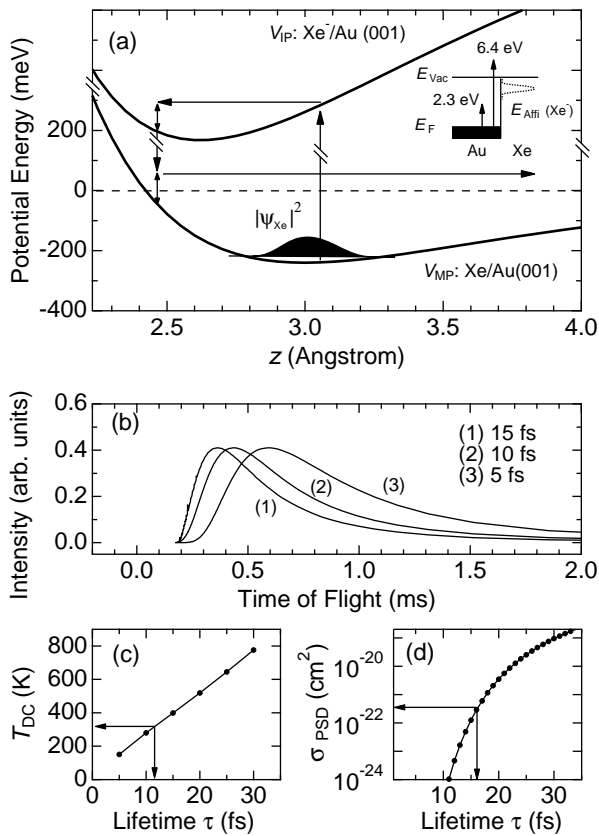} 
\caption{(a) Adiabatic potential of Xe and Xe$^{-}$ on Au(001) and a schematic of the Antoniewicz model of neutral desorption used for the model calculations. (b) Calculated results of time-of-flight (TOF) of desorbing Xe for several Xe$^{-}$ lifetimes. (c) Translational temperature ($T_{\mathrm{DC}}$) of the calculated TOF obtained by analyzing with a Maxwell-Boltzmann distribution as a function of Xe$^{-}$ lifetime. (d) Calculated result of the non-thermal PSD cross section as a function of Xe$^{-}$ lifetime. \label{ant}}
\end{center}
\end{figure}

The work function of the Au(001) surface is 5.0 eV, which is reduced by $\sim$0.5 eV with Xe adsorption. Therefore, the electronic states nearby the vacuum level are accessible with the hot electrons from the substrate band created by 6.4 eV photoexcitation, and not by 2.3 eV as schematically shown in Fig.~\ref{ant}(a). Although the electron affinity of Xe atoms in the gas phase is known to be negative,\cite{Buckman} the following studies suggest stabilization of the affinity level due to Xe condensation. Bulk Xe has a conduction band minimum (CBM) at 0.5 eV below the vacuum level.\cite{Schwentner1975, Schwentner1973} Haberland \textit{et al.} found that the ground-state Xe$_{N}$ clusters are able to bind an electron stably with $N>6$,\cite{Haberland} of which the electron affinity is calculated to be a few meV.\cite{Stampfli, Martyna} These studies suggest that interaction with neighboring atoms lowers the electron affinity level of Xe due to the mixing between the unoccupied orbitals. Furthermore, the image charge effect on metal surfaces shifts the electron affinity level downward by $\sim$1.0 eV.

Assuming that the excitation intermediate is the negative ion state, a plausible desorption mechanism is the Antoniewicz model.\cite{Antoniewicz} In the model, for the appreciable desorption to occur, the lifetime of the Xe$^{-}$ state is required to be long enough. We tentatively estimated the lifetime ($\tau$) of Xe$^{-}$ on Au(001) that reproduces the experimentally observed values of $T_{\mathrm{D1}}$ and $\sigma_{\mathrm{PSD}}$. $\sigma_{\mathrm{PSD}}$ is a product of the photoionization cross section $\sigma_{\mathrm{PI}}$ and desorption probability $P_{\mathrm{D}}$. We assume, as a first approximation, that the $\sigma_{\mathrm{PI}}$ is as large as $\sim$$10^{-16}$ cm$^{2}$.\cite{Moog} TOF of desorbing Xe and $P_{\mathrm{D}}$ are calculated on the basis of the Antoniewicz model and classical kinetics, as is depicted in Fig.~\ref{ant}(a). Initially, Xe atoms are trapped at the bottom of the physisorption well described by a Morse potential of the form $V_{\mathrm{MP}}(z)=D[1-\exp\{-\alpha(z-z_{0})\}]^{2}$, where $D$, $\alpha$ and $z_{0}$ represent the depth (240 meV),\cite{Mcelhiney} the width (14 nm$^{-1}$) and the position (3.0 \AA),\cite{Silva} respectively. The distribution of the initial position of Xe is accounted for as described in Ref.~\onlinecite{Moog}. Upon Xe$^{-}$ formation, the adiabatic potential of the Xe atom evolves into the form $V_{\mathrm{Ion}}(z)=V_{\mathrm{MP}}(z)+V_{\mathrm{IP}}(z)+\Delta E$, where $V_{\mathrm{IP}}(z)=-e^{2}/(16\pi\epsilon_{0}z)$ is the image charge potential and $\Delta E$ is an excitation energy. Due to the image charge attraction, the Xe atom is first attracted toward the surface, and is neutralized at a certain distance from the surface. The nuclear motion on $V_{\mathrm{Ion}}(z)$ is treated classically. If the Xe atom gains enough energy, it escapes the physisorption well leading to desorption. We assume that the neutralization rate of Xe$^{-}$ is described by $R(t)=\exp(-t/\tau)/\tau$ independent of $z$.

Figure~\ref{ant}(b) shows the TOF results of Xe calculated for lifetimes of 5$-$15 fs. Each TOF is well expressed by a single MB velocity distribution with a translational temperature $T_{\mathrm{DC}}$. Figure~\ref{ant}(c) shows the obtained $T_{\mathrm{DC}}$ of the calculated TOF as a function of $\tau$, where $\tau=$ $\sim$13 fs reproduces the experimentally observed $T_{\mathrm{D1}}$ of 300$\pm$20 K. Figure~\ref{ant}(d) shows the calculated result of $\sigma_{\mathrm{PSD}}$ as a function of $\tau$, where $\tau=17\pm5$ fs reproduces the experimentally observed $\sigma_{\mathrm{PSD}}$ of 10$^{-21}$$-$10$^{-22}$ cm$^{2}$. It is worth emphasizing that the two experimental data of $T_{\mathrm{D1}}$ and $\sigma_{\mathrm{PSD}}$ are well reproduced by a common $\tau$ value of $\sim$15 fs based on the Antoniewicz model. The fact is indicative of the validity of the present model. Walkup \textit{et al.} have shown that the classical adiabatic potential concerning the image charge potential is essentially correct, athough it is slightly different from the one obtained by a quantum-mechanical treatment. They have furthermore shown that the classical treatment of nuclear motion is valid as long as distribution of the initial Xe position is accounted for and that it qualitatively reproduces the kinetic-energy distribution.\cite{Walkup} It is noted that the affinity level of Xe should lie below the vacuum level for the $\tau$ to be as long as 15 fs.

The obtained value of $\tau$ $\sim$15 fs corresponds to the linewidth of 70$-$120 meV for the Xe$^{-}$ state. Padowitz \textit{et al.} found that in using the two-photon photoemission spectroscopy, the image charge state on clean Ag(111) is shifted by Xe adsorption due to the coupling with the Xe orbitals.\cite{Padowitz, Merry} The linewidth obtained in the present study is similar to the  value of 25$-$50 meV observed for the image charge states ($n=1, 2, 3$) on Xe/Ag(111) at 0.6$-$0.1 eV below the vacuum level. Hence, we suggest the image charge state of Xe/Au(001) is resonanced with the affinity level of Xe, which causes the PSD. We note that a smaller estimation of $\sigma_{\mathrm{PI}}$ $\sim$10$^{-18}$ cm$^{2}$ in the model calculation results in a linewidth of $\sim$35 meV.

As already mentioned above, two possible mechanisms of Xe$^{-}$ stabilization are hybridization of unoccupied orbitals and the image charge effect. Since unoccupied orbitals have an extended feature compared with occupied orbitals, unoccupied states could be appreciably hybridized with substrate states even in a weakly bound physisorption well. In addition to these two factors, we discuss another possible reason for the Xe$^{-}$ formation on a metal surface. In the gas phase, contrary to the ground state Xe (5$s^{2}$5$p^{6}$), metastable Xe$^{*}$ (5$s^{2}$5$p^{5}$6$s$) binds an electron to form a transient Xe$^{-}$ (5$s^{2}$5$p^{5}$6$s^{2}$) with a large cross section ($\sim$10$^{-16}$ cm$^{2}$).\cite{Blagoev} In the gas phase, the 5$s^{2}$5$p^{5}$6$s$ state is located at 8.3 eV above the ground state. A recent density functional study\cite{Silva} has shown that the Xe adsorption on a metal surface results in a partial depletion of the occupied Xe 5$p_{\mathrm{z}}$ state and a partial occupation of the previously unoccupied Xe 6$s$ and 5$d$ states. This indicates mixing of the 5$s^{2}$5$p^{5}$6$s$ state upon adsorption on a metal surface, which may contribute to the stabilization of the Xe$^{-}$ state. 

Lastly, we comment on the result of an earlier study on the nonthermal PSD of Xe from Ru(001) surfaces.\cite{Feulner1987} In the study, no significant desorption was observed from Xe mono and multilayers following 7$-$30 eV photoirradiations, whereas desorption from Ar mono- and multilayers and Kr multilayers were observed. As discussed in the present paper, Xe$^{-}$ is expected to be formed following the photoirradiations of $h\nu>$ 6 eV, and subsequently Xe desorption is expected to occur. Although the desorption cross section is not mentioned in Ref.~\onlinecite{Feulner1987}, we suspect $\sigma_{\mathrm{PSD}}$ of $\sim$10$^{-22}$ cm$^{2}$ was too small for the signal to be detected in their experimental condition. Arakawa \textit{et al.} reported that the absolute yield of the PSD from solid Ar following 12$-$50 eV photons is as large as $\sim$0.1 atoms/photon,\cite{Arakawa2} indicating that the cross section of the Xe PSD via the Xe$^{-}$ formation observed in the present study is several orders of magnitude smaller than those of Ar via exciton excitation.

\section{Conclusion}
In conclusion, we have investigated the PSD of Xe on Au(001) at photon energies of 2.3 and 6.4 eV.  With decreasing pump laser fluence, the desorption was found to undergo transition from thermal to non-thermal regimes.  The non-thermal PSD of Xe occurred only at 6.4 eV as a one-photon process, and the desorption proceeds via the Antoniewicz model with transient negative ion formation.  On the basis of the model calculation, the lifetime of Xe$ ^ { - } $ is estimated to be $\sim$15 fs. These results strongly suggest that the affinity level of Xe is substantially stabilized by the metal proximity effect.
 
\begin{acknowledgments}
This work was supported by Grant-in Aid for Scientific Research (A) of Japan Society for the Promotion of Science (JSPS) and by the Sasakawa Scientific Research Grant from the Japan Science Society. A. I. acknowledges support from a Research Assistant of the Global Centre of Excellence for Physical Science Frontier of Tokyo University, Japan.
\end{acknowledgments}

\bibliographystyle{ikeda}
\bibliography{./lidxe}

\end{document}